\documentstyle[aps,preprint,pre]{revtex}

\begin{document}
\tighten

\title{\bf 
Universality Class of Discrete Solid-on-Solid Limited Mobility
Nonequilibrium Growth Models for Kinetic Surface Roughening}

\author{S. Das Sarma$^{(1)}$, P. Punyindu Chatraphorn$^{(1,2)}$, 
and Z. Toroczkai$^{(1,3)}$}
\address{
$^{(1)}$Department of Physics, University of Maryland, College Park,
MD 20742-4111, USA \\
$^{(2)}$Department of Physics, Faculty of Science,
Chulalongkorn University,
Bangkok 10330, Thailand \\
$^{(3)}$Theoretical  Division and Center for Nonlinear Studies,
Los Alamos National Laboratory, Los Alamos, NM 87545, USA}

\date{\today}
\maketitle

\begin{abstract}
We investigate, using the noise reduction technique, the
{\it asymptotic} universality class of the well-studied nonequilibrium
limited mobility atomistic solid-on-solid surface growth models
introduced by Wolf and Villain (WV) and Das Sarma and Tamborenea (DT)
in the context of kinetic surface roughening in ideal molecular beam
epitaxy. We find essentially all the earlier conclusions
regarding the universality class of DT and WV models to be severely
hampered by slow crossover and extremely long lived transient
effects. We identify the correct asymptotic universality class(es)
which differs from earlier conclusions in several instances.
\end{abstract}

\vskip 1pc
\vfill\eject

Kinetic surface roughening of nonequilibrium growth models, particularly
under solid-on-solid (SOS) growth conditions, remains a subject of 
considerable interest and activity in spite of a great deal of theoretical
and experimental research in the topic during the last decade
\cite{intro}.
While the {\it generic} nonequilibrium surface growth universality class
(in the situation allowing overhangs and bulk defects, i.e. under
generic non-SOS conditions) is theoretically accepted to be the KPZ
universality (although the asymptotic KPZ universality in specific
models may be masked by extremely slow crossover effects), there
is no such consensus for SOS growth models where seemingly innocuous
small changes in local growth rules appear to lead to different dynamic
universality classes.
In fact, some of these nonequilibrium SOS growth models, introduced
in the context of mimicking ideal molecular beam epitaxy (MBE),
do not even seem to obey self-affine dynamic scaling behavior,
instead exhibiting nonuniversal anomalous and multi-affine scaling.
It may therefore be questioned whether the universality class
concept is useful in SOS growth models or even applies to nonequilibrium
SOS growth models.

In this paper we address this confusing situation plaguing our 
understanding of the universality class of discrete SOS nonequilibrium 
growth models by concentrating on a specific class of surface 
diffusion driven stochastic growth models which mimic in a drastically
simplified manner, low temperature molecular beam epitaxial growth.
This class of models has been referred to as `limited mobility'
nonequilibrium growth models because the deposited atoms are only
allowed to diffuse (obeying certain specific local diffusion rules)
at incidence, and all the other atoms in the growing film, except
for the most recently incident atom, do not diffuse. 
In spite of their highly simplified nature, limited mobility 
nonequilibrium surface growth models have attracted a great deal of
attention \cite{intro}
primarily because of the following three reasons:
(1) these are the first growth models which were demonstrated to
lie outside the generic universality class;
(2) in spite of their highly simplified nature, the various growth
exponents and growth morphologies in these models seem to agree well
with those in full diffusion Arrhenius hopping MBE growth models
(where all atoms at the growth front are allowed to hop continuously
according to temperature dependent hopping rates) in the low
temperature kinetically rough regime;
(3) obtaining a coarse-grained {\it continuum} description of the
{\it discrete} low mobility growth models (i.e. writing down a 
continuum dynamical growth equation corresponding to the cellular
automata discrete rules defining the low mobility growth models)
has turned out to be extremely difficult in spite of the deceptive
simplicity of the models.
The innovation and the new technique we introduce in the study of
these extremely well-studied limited mobility growth models is the
use of the {\it noise reduction} technique 
\cite{ppnrt} in the direct numerical
simulation of the atomistic growth rules.
Our conclusion, based on extensive numerical simulations in both
1+1 and 2+1 dimensions (plus some limited simulations in higher
dimensions) is that the {\it asymptotic} universality class of the
various limited mobility growth models is a surprisingly subtle
issue with many of the earlier findings (including some from our 
group) are {\it incorrect} due to pathologically slow crossover
and extremely long transient effects which typically distract from
ascertaining the true asymptotic universality class of these
models. 

One strikingly novel feature of our results is the apparent
dependence of the asymptotic universality class on the system
dimensionality ($d$), not merely in the sense of well-known 
hyperscaling relations, but in the fact that the applicable
hyperscaling relation itself (connecting the dynamical exponent $z$ 
with the roughness exponent $\alpha$ for example) for a specific 
limited mobility growth model may depend on $d$.
We believe that such dimension-dependent {\it nonuniversal} 
universality, where the hyperscaling relation for a given model
changes with $d$ (and which, to the best of our knowledge, has no
known analog in equilibrium critical phenomena where a given 
hamiltonian or free energy functional, e.g. the $\phi^4$-model,
is characterized by a unique hyperscaling relation which
{\it does not} change with the specific value of $d$),
is a characteristic feature of nonequilibrium processes, and
may transcend the specific models and phenomena being studied here.
One implication of this peculiar dimension-dependent universality
is that the universality class concept, which is one of the most 
important conceptual foundations of modern critical phenomena,
may be of rather limited validity and usefulness in nonequilibrium
phenomena since the model by itself does not specify the 
universality class --- it depends on the model (i.e. the specific
set of discrete dynamic growth rules in our growth model)
{\it and also on the dimensionality}.

We study the limited mobility nonequilibrium growth models 
introduced by Wolf and Villain (WV) \cite{wv1}
and by Das Sarma and Tamborenea (DT) \cite{dt1},
both of which have already been studied extensively in the
literature and discussed rather elaborately in recent reviews
\cite{dt2}.
We also study two simple variants of these models, which we call
{\it asymmetric} DT (ADT) and {\it symmetric} WV (SWV) models.
The growth rules for these models are described in Fig. \ref{rule}
for $d=1+1$ --- the rules for higher dimensions involve 
straightforward generalization of the rules depicted in 
Fig. \ref{rule}. In all four models atoms are deposited randomly
on a ($d-1$)-dimensional square lattice substrate
(which is flat initially) obeying the SOS constraint ---
a deposited atom is then allowed to diffuse or hop instantaneously
at incidence
(all limited mobility models are also by definition,
{\it instantaneous relaxation} models) to its final incorporation
site (after which the {\it incorporated} atom never moves again)
according to the bonding configurations of the deposition site.
In DT(WV) model the deposited atom tries to increase(maximize)
its nearest-neighbor bonding configuration by moving to the
incorporation site. If the deposition and all possible 
incorporation sites have the same nearest-neighbor bonding
configurations, the incident atom does not move and stays at the
site of deposition. The diffusion length ($l$), the distance 
over which the deposited atom is allowed to search to find its
incorporation site, is a parameter of the model, and the 
simulation results shown in this paper use $l=1$ (with the 
length unit being the lattice constant throughout this paper),
i.e. nearest-neighbor diffusion only.
We have, however, verified that $l$ is an irrelevant variable, 
and all our conclusions in this paper are independent of the precise
value of $l$ as long as $l \ll L$, where $L$ is the linear size of 
the substrate, although finite size and crossover effects may be 
strongly (and nontrivially) dependent on the value of $l/L$.
The DT model has the additional diffusion constraint that only
deposited atoms with no lateral nearest-neighbor bonds (i.e. the
random incidence site has no occupied nearest-neighbor in the same
``layer'') are allowed to move whereas in the WV model {\it all}
deposited atoms may move provided the local coordination number
is maximized. The ADT and the SWV models are intermediate in their 
dynamics with respect to DT and WV models in the sense that ADT 
maximizes the local coordination while still allowing only deposited
atoms with no nearest-neighbor lateral bonds to diffuse whereas
the SWV model only increases the local coordination while allowing
all deposited atoms to move (provided they can increase their
local coordination or nearest neighbor bonding configuration).
In all the models, the incident atom is allowed to move randomly
to the incorporation site if several possible incorporation sites 
satisfy the growth rules. 

All our simulations utilize the noise reduction technique introduced 
by two of us in an earlier paper \cite{ppnrt}. 
The noise reduction technique allows
only a fraction of the successful hits (i.e. the incident atoms
which satisfy the specific growth rules of the model) to be executed 
--- the noise reduction parameter $m$ defines the number of successful
hits needed before incorporation is allowed ($m=1$ is the original
DT or WV model). It is well-known that the noise reduction technique,
which has been extremely successful in clarifying the universality 
class of various non-SOS growth models (e.g. Eden model, DLA)
with severe crossover problems, is very effective in obtaining 
the universality class of growth models. In particular, the noise
reduction technique ($m>1$) effectively suppresses the severe 
correction to scaling induced by the strong stochasticity in
limited mobility growth models, and thus makes it easier to
determine the asymptotic universality class. We emphasize that
the noise reduction technique is absolutely crucial in obtaining the
results and conclusions presented in this paper, and its introduction
in the simulations is the key feature which enables us to determine
the universality class(es) of the limited mobility growth models
we study in this paper.

Before presenting our simulation results we briefly describe the
continuum dynamical growth equation approach we use in discussing
the universality class of various models. Denoting the dynamical
height fluctuation variable as $h$, and the lateral coordinate
(along the substrate) as ${\bf x}$ and the growth ``time'' as $t$
(note that the growth time is defined entirely by the average 
deposition rate since we neglect bulk vacancies, overhangs, and
evaporation in our SOS model) the most general leading order growth
equation for our problem can be written as
\begin{equation}
\frac{\partial h}{\partial t} = \nu_2 \nabla^2h - \nu_4
\nabla^4h + \lambda_{22} \nabla^2
( \mbox{\boldmath $\nabla$} h )^2 + \eta,
\label{eq4th}
\end{equation}
where $h \equiv h({\bf x},t)$, and $\eta$ is the spatio-temporal
Gaussian white noise associated with the incident beam shot noise.
The symmetry-allowed fourth order nonlinear term 
$\mbox{\boldmath $\nabla$}(\mbox{\boldmath $\nabla$} h)^3$
has been left out of Eq. (\ref{eq4th}) since, upon renormalization,
it generates the linear $\nabla^2 h$ term already included in
Eq. (\ref{eq4th}). (The constant flux term associated with the
average growth has been left out of Eq. (\ref{eq4th}) since the
height fluctuation variable $h$ is defined with respect to the
average interface position.) When $\nu_2 \neq 0$, the asymptotic
growth universality class is the so-called EW universality
($\nu_2>0$) or the ``unstable'' or mounded growth universality
($\nu_2<0$) --- the fourth order terms in Eq. (\ref{eq4th})
are then all irrelevant although they may be important in controlling
the non-asymptotic transient regime which may last for a long time.
When $\nu_2 \equiv 0$ (as well as the corresponding fourth order 
nonlinear term 
$\mbox{\boldmath $\nabla$}(\mbox{\boldmath $\nabla$} h)^3$
being absent), the asymptotic growth universality class is the 
MBE growth universality determined by the 
$\nabla^2 ( \mbox{\boldmath $\nabla$} h )^2$ term 
with the initial transient regime (which could be extremely long-lived
depending on the ratio $\nu_4/\lambda_{22}$) controlled by the 
irrelevant fourth order linear term.
When both $\nu_2=0$ and $\lambda_{22}=0$, the growth universality 
class is the MH universality defined by the $\nu_4 \nabla^4h$ term.
The various growth exponents for these universality classes are
well-known and can be found in the literature.

Determining the asymptotic growth universality class of a discrete 
nonequilibrium growth model may be difficult due to extremely
long-lived transients which would mask the crossover behavior.
This is obviously a more acute problem when any two (or all three)
of the growth terms ($\nu_2$, $\lambda_{22}$, $\nu_4 \neq 0$)
are nonzero in Eq. (\ref{eq4th}) although complications could 
also arise from higher order terms (i.e. higher than $4$th order)
neglected in Eq. (\ref{eq4th}). It is reasonable to assume that in 
the generic situation, i.e. in the absence of any compelling 
symmetry or conservation law induced constraints, the fourth order
nonlinear equation, Eq. (\ref{eq4th}), should suffice to determine
the asymptotic universality class of a given growth model, and
there is no need to consider a higher-order (e.g. the sixth order)
dynamical equation.
This is particularly true since simple power counting considerations
indicate that many of the higher order nonlinearities produce the
$\nu_2$ or the $\lambda_{22}$ term upon renormalization,
and most of the higher order terms are simply irrelevant.
Thus it is quite possible that even in the extremely unlikely 
situation that all the terms in Eq. (\ref{eq4th}) vanish, i.e.
$\nu_2$, $\lambda_{22}$, $\nu_4 = 0$, for some pathological reasons
Eq. (\ref{eq4th}) may still remain a valid generic description of the
asymptotic growth universality class, because higher order nonlinearities
neglected in Eq. (\ref{eq4th}) give rise to the terms
$\nabla^2h$ and/or 
$\nabla^2(\mbox{\boldmath $\nabla$}h)^2$
in the growth dynamics.

We have determined the asymptotic universality class from our 
discrete stochastic simulations by calculating the growth exponents
characterizing the height-height correlation functions as well as
by measuring the surface current on a tilted substrate as
proposed in ref. \cite{krugj}.
We emphasize that the noise reduction technique is crucial in
enabling us to conclusively determine the asymptotic universality
class of the four (DT, WV, ADT, SWV) discrete growth models we
study in contrast to earlier simulational studies which have been
severely hampered by slow crossover, long transient, and strong
correction to scaling problems. Our conclusions about the growth
universality classes of the four models, as summarized in
Table 1, are based on consistent results from simultaneous
measurements of growth exponents and surface currents 
on tilted substrates.

The result in $d=1+1$ for the DT and the WV model are already known 
in the literature \cite{ppnrt,wv1,dt2,krugj}
from extensive earlier one dimensional simulations
of these two well-studied models.
The one dimensional DT (WV) model belongs to MBE (EW) asymptotic
universality as we also confirm decisively in our current 
simulations. 
The fact that the $d=1+1$ DT model has $\nu_2 = 0$ 
exactly follows from a hidden dynamical symmetry 
\cite{kruglong} in the DT
growth rules which are {\it symmetric}, i.e., the DT diffusion
rules do not have any preference between two- and three-
bonded incorporation sites. This also makes the current on a tilted
substrate exactly vanish in the $d=1+1$ DT model, confirming that
it is generically in the MBE universality class
[$\nu_2 = 0$ in Eq. (\ref{eq4th})].
Similarly, earlier tilted substrate current measurements 
\cite{krugj} as well
as direct simulational exponent measurements 
for the $d=1+1$ 
WV model establish it to be in the EW universality class
(i.e. $\nu_2 > 0$) although the WV model shows very similar
scaling behavior to the DT model for a very long transient
regime since $|\nu_2| \approx 0$ and $\nu_4$, $\lambda_{22}$
are rather large in the WV model.
Our noise reduced simulations of the WV model 
\cite{ppnrt} verify rather
strikingly that this model belongs to EW universality in
$1+1$ dimensions.
The suppression of crossover and correction to scaling effect
in the $1+1$- dimensional DT and WV models was our original 
motivation for introducing \cite{ppnrt}
the noise reduction technique 
in this context.

The kinetically rough surface growth is traditionally analyzed     
in terms of the dynamic scaling hypothesis where the surface 
width (the root mean square fluctuation in the surface height)
or more generally, the height-height correlation function
shows generically scale invariant power law scaling behavior,
with critical exponents $\alpha$, $\beta$, $z=\alpha/\beta$,
given by 
$W(L,t) \sim L^\alpha f (\xi(t)/L)$
where $W$ is the dynamical surface width at growth time $t$
for a substrate of lateral size $L$ and $\xi(t)$ is the
lateral dynamical correlation length for the specific growth
process with $\xi(t) \sim t^{1/z}$.
The scaling function $f(x)$ behaves as 
$f(x \gg 1) \sim 1$ and $f(x \ll 1) \sim x^\alpha$
so that $W(L, t \rightarrow \infty) \equiv W_s(L) \sim L^\alpha$
and $W(L, t \ll L^{1/z}) \sim t^\beta$
with $\beta = \alpha/z$, where $W_s(L)$ is the saturated 
steady-state long time surface width and 
$W(L \ll \xi(t))$ is the pre-steady-state dynamical surface width.
The set of critical exponents $\alpha$, $\beta$, and $z=\alpha/\beta$ 
define the growth universality class, and in principle,
can be determined for a particular continuum growth equation.
For the sake of completeness (see the WV growth results
presented below) we point out that for mounding instability
with slope selection, when the surface morphology evolves into 
a regular mounded pattern with the sides of the mounds having
constant slopes, one gets $\alpha \equiv 1$, and only one exponent
($\beta$ or $z$) $\beta = 1/z$ then defines the growth pattern.
In Table 1 and in Figs. \ref{dtfig} and \ref{wvfig} 
we present our calculated critical
exponents for the DT and WV model along with the associated 
continuum equation descriptions.

Our most surprising findings are presented in 
Fig. \ref{dtfig} (for the DT model)
and Fig. \ref{wvfig} (WV model) where our results for the 
$2+1$- dimensional DT and WV models are depicted.
We find that the DT (WV) model in $2+1$ dimensions
belongs to the EW (unstable) dynamic universality in
contrast to the $1+1$ dimensional universality class
of these models.
This is the important new result presented in this paper
which disagrees with the earlier conclusions in the literature.
The determination of the asymptotic universality class of the
$2+1$- dimensional DT and WV model is the main new result
being presented in this paper.

We first discuss the $2+1$ dimensional DT results which are
in some sense less surprising than the corresponding WV
results because EW universality is the generic SOS universality
class.
Our measurement of current on tilted substrates in the 
$2+1$ dimensional noise reduced DT simulations always 
exhibit a small but finite downhill current indicating
the presence of a small $\nu_2 > 0$ term in Eq. (\ref{eq4th}).
This should be contrasted with the corresponding $1+1$- dimensional
DT results where a simple symmetry argument as well as 
extensive numerical simulations definitively establish
the absence of the $\nabla^2 h$ term in the $d=1+1$
DT growth equation, i.e. $\nu_2 = 0 (\neq 0)$ in the
$d=1+1(2+1)$ DT model.
The asymptotic universality class of $1+1$- dimensional
DT model is now well-established to be given by the following
continuum growth equation, which does {\it not} have the
generic $\nu_2 \frac{\partial^2 h}{\partial x^2}$ term
of Eq. (\ref{eq4th}):
\begin{equation}
\frac{\partial h}{\partial t} = 
\nu_4 \frac{\partial^4 h}{\partial x^4}
+\lambda_{22} \frac{\partial^2}{\partial x^2}
 (\frac{\partial h}{\partial x})^2
+\sum_{n=4,6,...} \lambda_{2n} \frac{\partial^2}{\partial x^2}
 (\frac{\partial h}{\partial x})^n
+\eta,
\end{equation}
where the ``higher order'' terms of the form
$\frac{\partial^2}{\partial x^2}
(\frac{\partial h}{\partial x})^{2m}$
with $2m \equiv n = 4,6,8,...$, etc.
are marginally relevant in $d=1+1$ (i.e. simple power counting 
reveals them as having the same anomalous dimensions as the
nonlinear fourth order $\lambda_{22}$ term).
Our tilted substrate current measurement in $d=2+1$ DT model
shows the existence of a small slope dependent downhill current
of appoximate magnitude $\sim 10^{-2}$ whereas the 
corresponding $d=1+1$ DT model has a current $\sim 10^{-6}$
(of random sign) which is indistinguishable from the 
background noise effect. We therefore conclude that
the $2+1$ dimensional DT model, describing nonequilibrium
growth on physical surfaces and interfaces, belongs to the
generic EW universality class, and has the following 
coarse-grained continuum description:
\begin{equation}
\frac{\partial h}{\partial t} =
\nu_2 \nabla^2 h - \nu_4 \nabla^4 h
+ \sum_{n=1,2,3,...} \lambda_{2(2n)} \nabla^2
(\mbox{\boldmath $\nabla$}h)^{2n} + \eta.
\label{eqdtfull}
\end{equation}
Our finding that $\nu_2$ is very small, but nonzero,
in Eq. (\ref{eqdtfull}) for $2+1$- dimensional DT growth
should not come as a big surprise (except, of course,
for the fact that it has not earlier been discovered in
the literature including our own earlier work on the
DT model) \cite{dtpp}
because the vanishing of $\nu_2$ in the $1+1$- dimensional
DT model arises from a rather peculiar kinetic-topological 
symmetry of the DT model, which applies only in one dimension
and cannot be generalized to two dimensional surfaces.
In the absence of a compelling symmetry argument manifestly
making $\nu_2=0$ in the growth equation, one should expect
its generic presence in the $2+1$- dimensional DT model
although the extreme quantitative smallness of $\nu_2$
has made it difficult so far to establish its finiteness
in simulations.
Our finding that there is a small downhill current on tilted
substrates in the $2+1$- dimensional DT model and that
the critical exponents of $2+1$- dimensional DT growth
(Fig. \ref{dtfig}) are consistent with EW universality class
(and {\it not} particularly consistent with the MBE 
universality defined by $\nu_2 = 0$ in Eq. (\ref{eq4th}))
leads to the conclusion that $2+1$- dimensional DT growth 
is in the EW universality class ($\nu_2 \neq 0$ in
Eq. (\ref{eq4th})) and $1+1$- dimensional DT growth is
in the MBE universality class ($\nu_2 = 0$).
We have also carried out DT simulations in (unphysical)
$3+1$ dimensions finding very good agreement with EW 
universality properties.

Our results for the $2+1$- dimensional WV model (Fig. \ref{wvfig})
are very dramatic and completely unanticipated. We find that
the $2+1$- dimensional WV model leads to spectacular        
quasiregular mounded morphology indicating unstable
epitaxial growth (whereas the corresponding $1+1$- dimensional
WV growth is asymptotically stable and flat, belonging to
the EW universality class). 
Thus the usual critical exponents ($\alpha$, $\beta$, 
$z=\alpha/\beta$) are not particularly meaningful for
$2+1$- dimensional WV growth (although they can still be
defined in the simulation results --- the exponents, however,
provide a misleading picture since the growth front, instead of
being statistically scale invariant as it should be for kinetic
surface roughening exhibiting power laws controlled by 
critical exponents, has a quasiregular mounded pattern).

The tilted substrate current measurement in the $2+1$-
dimensional WV growth yields another curious surprise.
It turns out that the local current is uphill (along 111 plane)
or donwhill (along 100) depending on how (i.e. along which
direction) one decides to tilt. The fact that the tilted
substrate current could depend sensitively (i.e. stabilizing
in some directions and destabilizing in other directions)
on the tilt direction has not earlier been reported in the
literature where most reported current measurements are
carried out in $1+1$ dimensions (where, of course, this problem
cannot arise) with the hope or expectation (proven to be false
in this paper) that an accurate determination of the 
universality class of a growth model in $1+1$ dimensions
will automatically give us the {\it same} universality class
in $2+1$ dimensions --- WV model is in the EW universality
class in $1+1$ dimensions and unstable (mounded morphology)
in $2+1$ dimensions!
Thus the tilted substrate current measurement, while being
capable of providing the correct universality class in $1+1$
dimensions, may very well lead to misleading and wrong conclusions
in higher dimensions where the current on tilted substrates is
explicitly direction dependent and is not uniquely defined.
We will publish quantitative details on the WV mounding 
phenomenon elsewhere --- here we point out that
(1) the underlying mechanism for the WV mounded morphology
is related to surface cluster-edge (or kink) diffusion
induced mounding recently discussed in the literature
\cite{sedref},
and (2) this WV mounding phenomenon, arising as it does 
from kinetic-topological aspect of surface diffusion,
leads to very strong instabilities in (unphysical)
dimensions higher than $2+1$, where early work reported \cite{wv3d}
unexplained strong mound formation in $3+1$ and $4+1$ 
dimensional WV growth.
We have carried out WV simulations in $3+1$ dimensions,
finding very strong mounding even without any noise reduction
consistent with earlier findings \cite{wv3d}.

Finally we consider the two intermediate models, ADT and
SWV, for the sake of completeness. 
The d=2+1 growth morphologies in these two models are 
shown in Figs. \ref{adtfig} (ADT) and \ref{swvfig} (SWV).
Without presenting the actual numerical results for
the critical exponents for these
two models, we just mention that the ADT and SWV models
in 2+1 dimensions
behave qualitatively similar to the WV model in $2+1$
dimensions with fairly strong mounding under noise reduction
although the morphological details in the two models
differ somewhat with the SWV morphology being similar to
the pyramidal structures of the WV morphology and the ADT
morphology having flat top mounds with very deep and narrow
grooves. 
Thus, ADT/SWV/WV all have unstable growth in
$2+1$ (or higher) dimensions with quasiregular mounded
morphology whereas DT in $2+1$ (or higher) dimensions
is in the EW universality with stable and smooth growth
morphology. All four models have kinetically rough 
statistically scale invariant growth in $1+1$ dimensions
with the WV (DT) model belonging to EW (MBE)
universality class.

We conclude by stating that we have found the universality class
concept to be of limited usefulness in conserved discrete 
limited mobility nonequilibrium surface growth models.
The same growth rules defining a particular model (e.g. WV
or DT) may belong to different universality classes in
different dimensionalities (not in the sense of 
super-universality, but in a more fundamental nontrivial 
sense as if an equilibrium model, which is in the Ising
universality class in two dimensions, behaves as an
x-y model in three dimensions --- a patently absurd notion).
In addition, rather minor changes in local growth rules could 
lead to dramatic differences in the resulting growth morphology
or the universality class --- DT and WV have very similar 
local growth rules, but their morphologies, smooth (DT) and
mounded (WV), and universality class (EW for DT and unstable
for WV) are strikingly qualitatively different. We also 
find that measuring surface current on tilted substrates 
\cite{krugj}, 
while being a potentially useful technique for discussing the
$1+1$ dimensional universality class, may not work in
$2+1$ dimensions and may produce misleading or conflicting
conclusions depending on the precise direction of the
surface current.             
We have assumed throughout that the noise reduction technique,
which is absolutely crucial in our obtaining the asymptotic
universality classes of various models we study, does not
modify the universality class of a growth model
(and only suppresses transient and correction-to-scaling effects 
by reducing the effective noise strength).
This belief is based on extensive earlier analysis of the 
noise reduction technique in the literature, which,
in general, is thought not to affect the growth universality class.

This work is supported by the NSF-MRSEC and US-ONR.

\begin{figure} \caption{
Schematic configuartions defining the growth rules for the
DT and WV models in $d=1+1$. For ADT model, adatom "g" in
the top (DT) plot will only hop to the left where it will have
3 nearest neighbors. For SWV model, adatom "g" in the bottom
(WV) plot will choose between the left and right neighbors
with equal probability.
\label{rule}} \end{figure}

\begin{figure} \caption{
(a) The interface width $W$ versus time $t$ in 2+1 DT model with
substrate size $L=1000 \times 1000$ and noise reduction factor
$m$ = 1, 3, 10, and 15 from top to bottom. Solid lines are best
power law fit which yield the growth exponent $\beta$ that decreases
as $m$ increases.
(b) A typical morphology of a noise reduced DT model ($m=5$)
from a $1000 \times 1000$ substrate (only a section of 
$200 \times 200$ is shown here) at 400 ML. 
\label{dtfig}} \end{figure}

\begin{figure} \caption{
(a) The interface width $W$ versus time $t$ in 2+1 WV model with
substrate size $L=100 \times 100$ and noise reduction factor
$m=5$. 
(b) A typical morphology of a noise reduced WV model ($m=5$)
from a $500 \times 500$ substrate (only a section of 
$200 \times 200$ is shown here) at $10^6$ ML.
\label{wvfig}} \end{figure}

\begin{figure} \caption{
A typical morphology of a noise reduced ADT model ($m=5$)
from a $100 \times 100$ substrate at $10^6$ ML.
\label{adtfig}} \end{figure}

\begin{figure} \caption{
A typical morphology of a noise reduced SWV model ($m=5$)
from a $100 \times 100$ substrate at $10^6$ ML.
\label{swvfig}} \end{figure}

\begin{table}[!h]
\centering
\begin{tabular}{|c|c|c|c|}  \hline
model & properties & d=1+1 & d=2+1 \\  \hline
\, & exponents & $\alpha=1$, $\beta=1/3$, $z=3$ &
                 $\alpha=0(\log)$, $\beta=0(\log)$, $z=2$ \\
                 \cline{2-4}
DT & morphology & kinetically rough & flat and very smooth \\  \cline{2-4}
\, & particle current & $J=0 (\sim \pm 10^{-6})$ &
                        $J \sim -10^{-4}$ \\  \hline
\, & exponents & $\alpha=1/2$, $\beta=1/4$, $z=2$ &
                 $\alpha=1$, $\beta=1/3$, $z=4$ \\
                 \cline{2-4} 
WV & morphology & kinetically rough & mounds with selected slope
                 \\  \cline{2-4}
\, & particle current & $J \sim -10^{-3}$ &
        $J_{100} \sim -10^{-2}$; $J_{111} \sim +10^{-2}$ \\  \hline
\end{tabular}
\vskip 1cm
\caption{
Summarized of our results for the DT and WV models
}
\label{tab1}
\end{table}
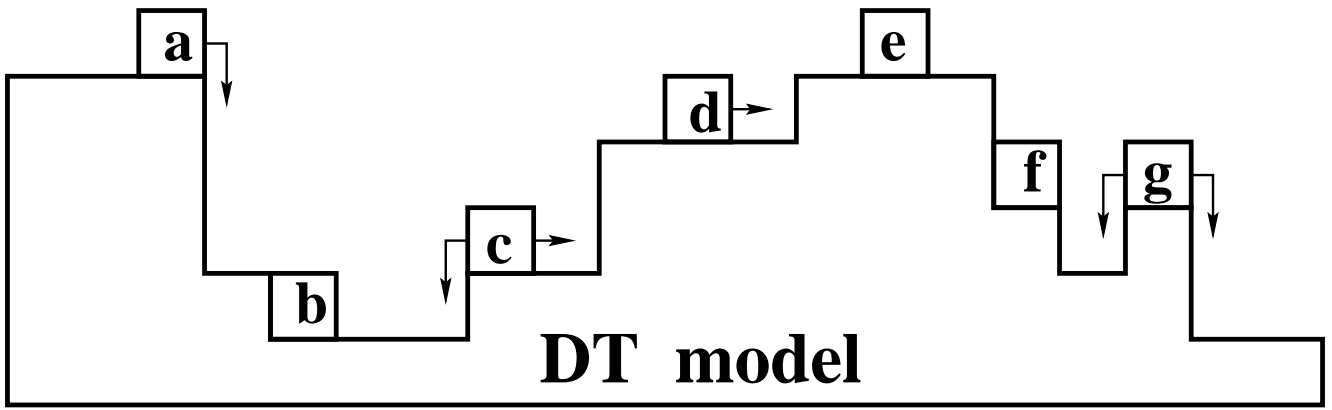
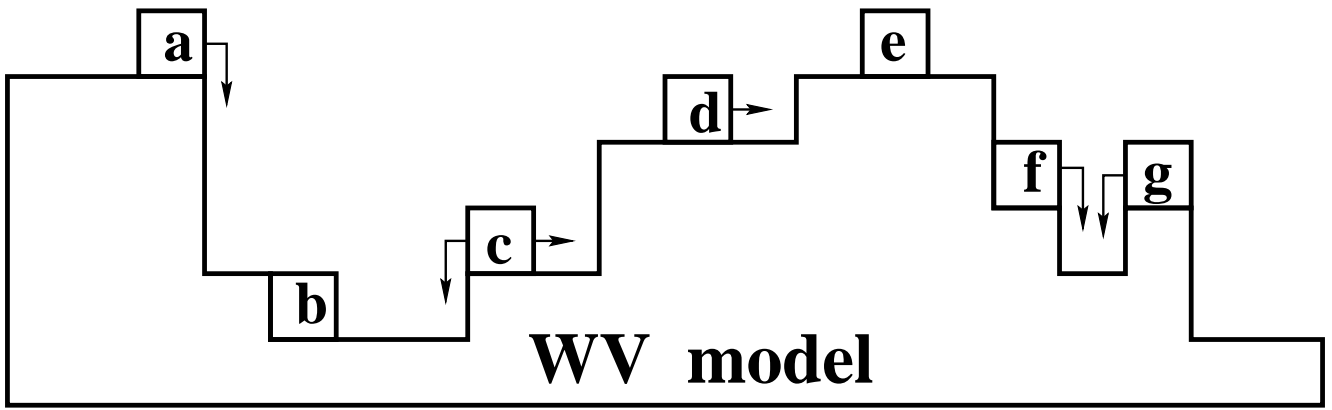
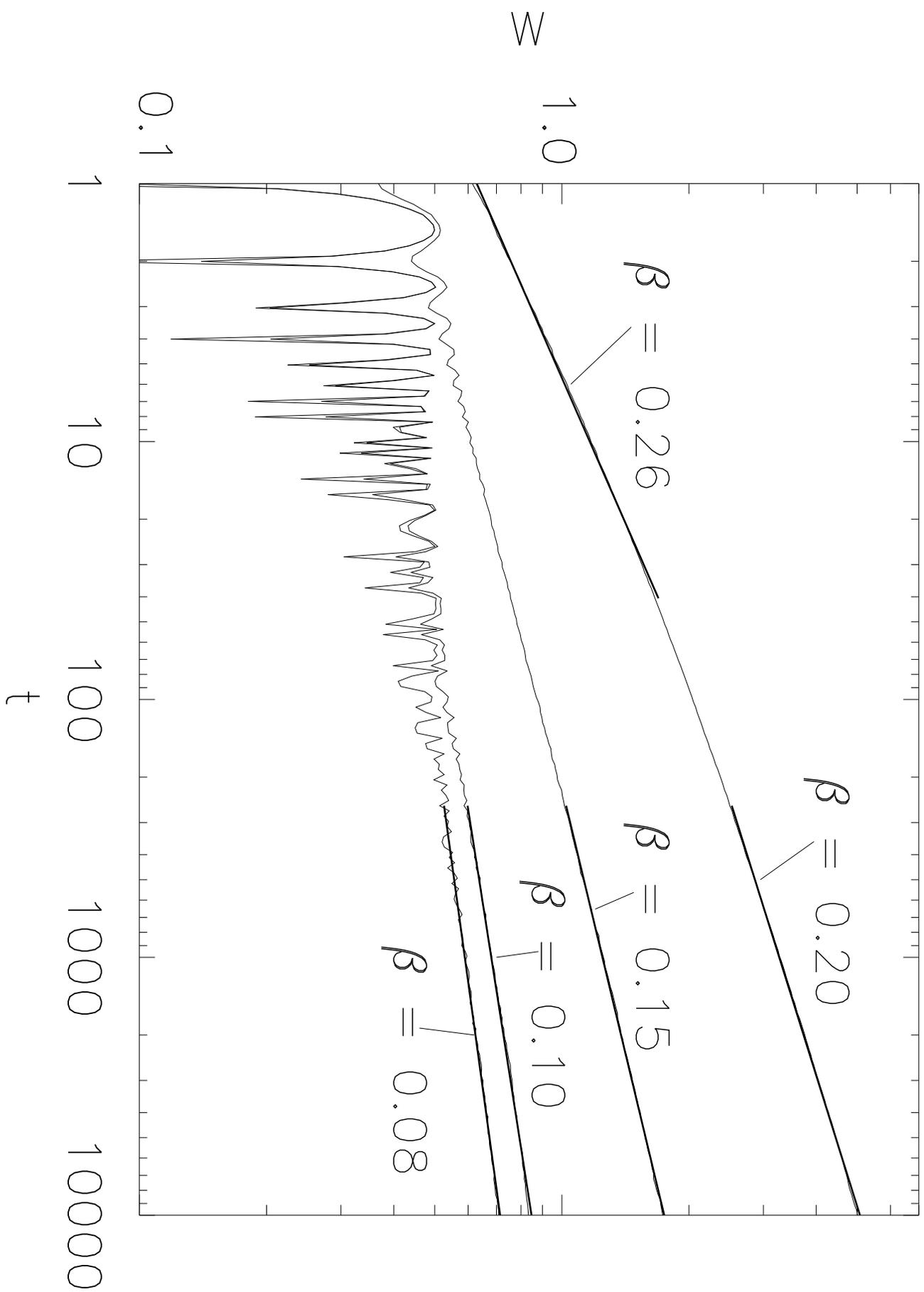
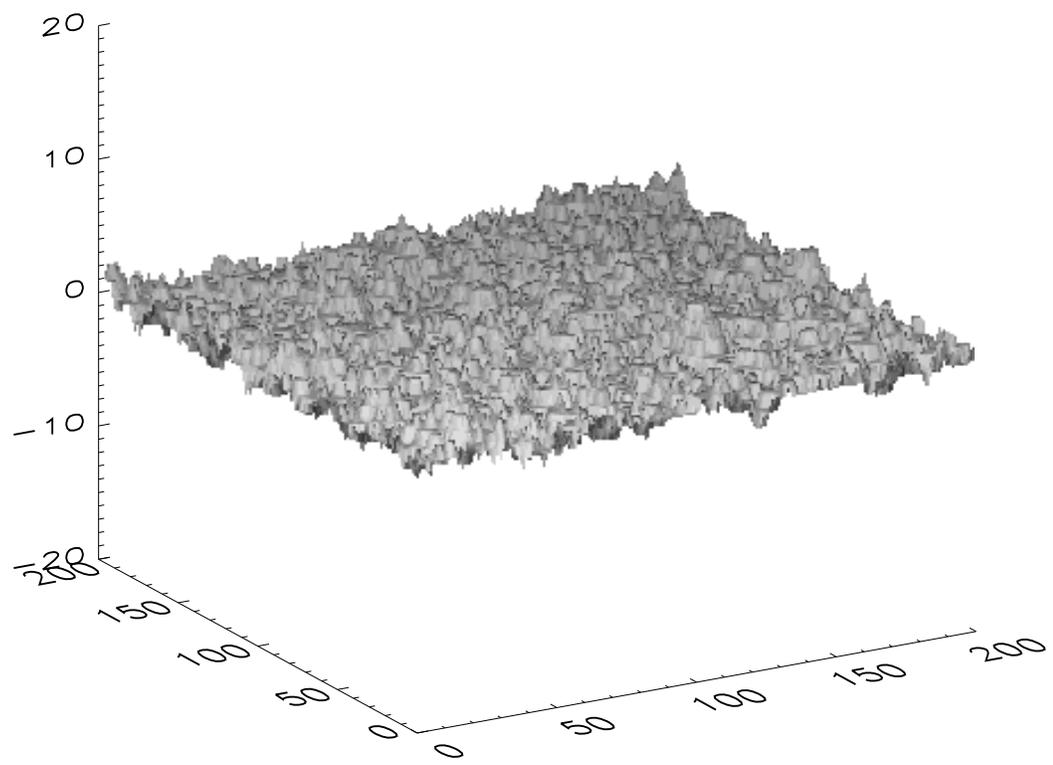
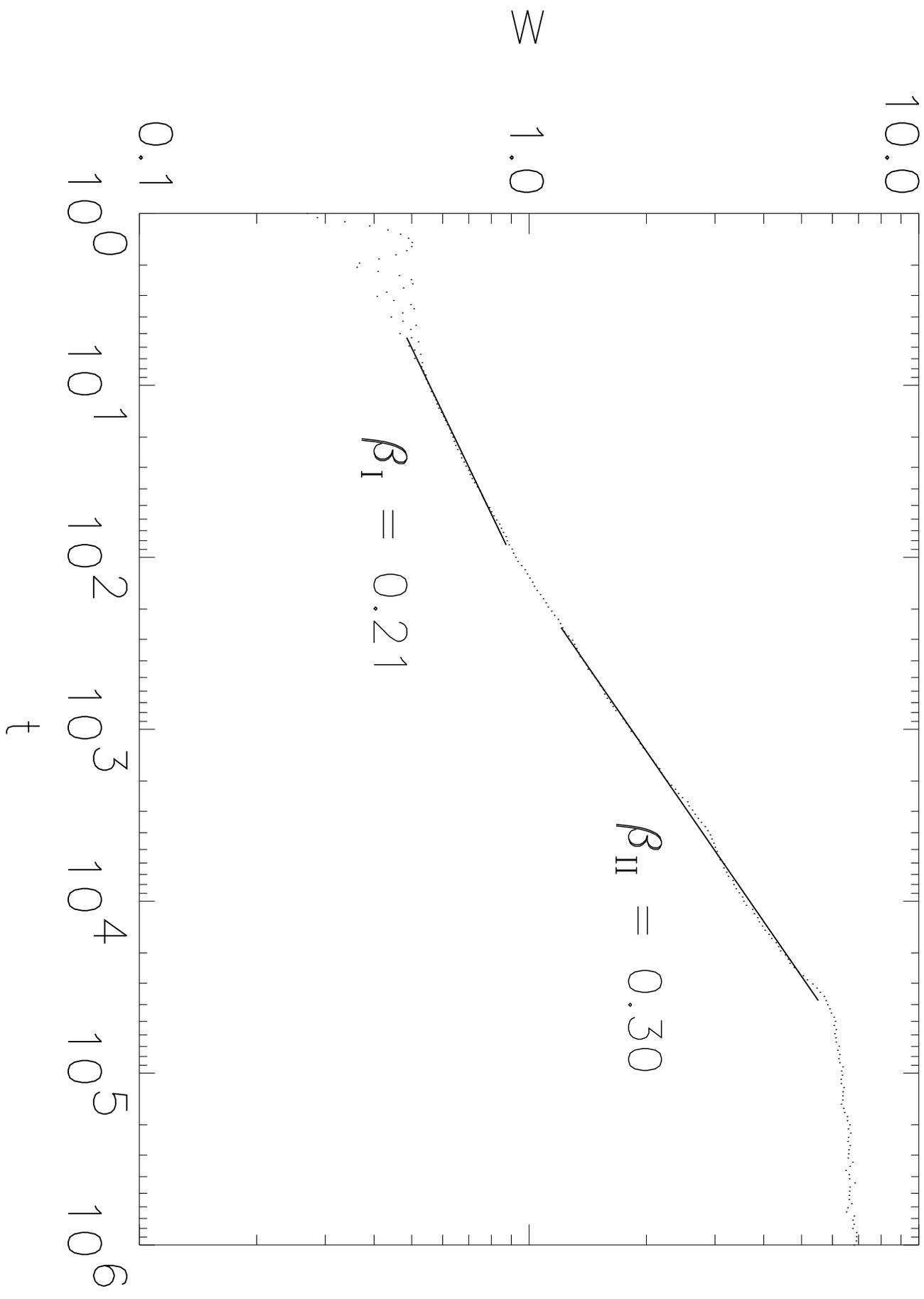
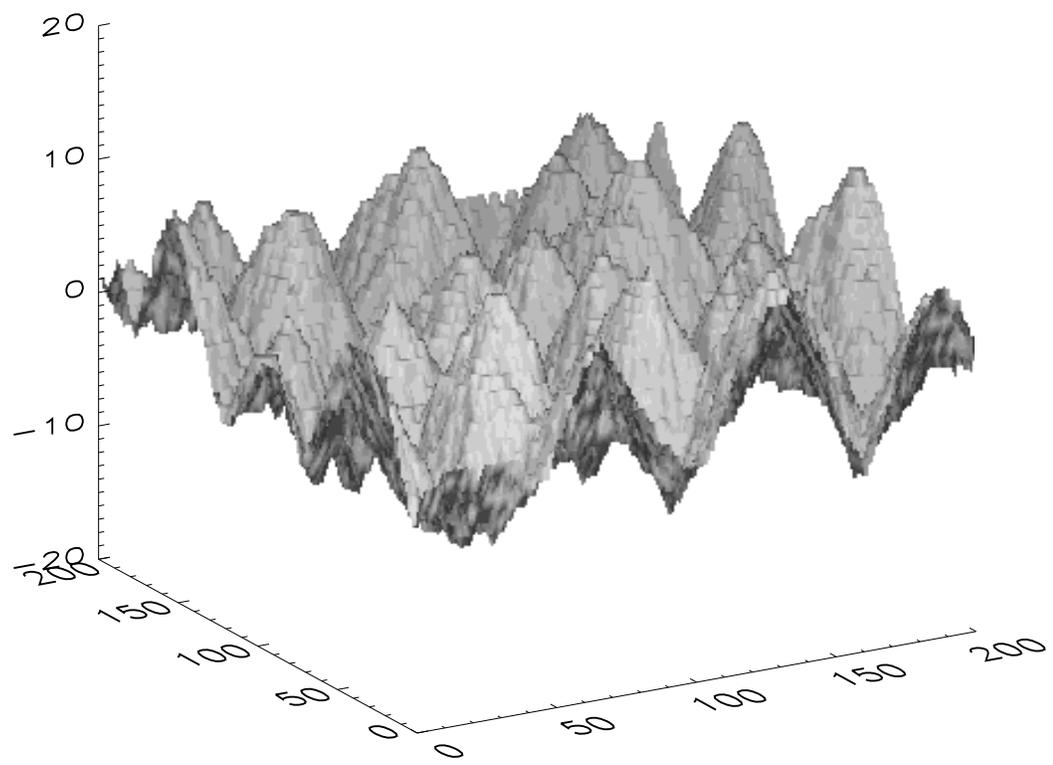
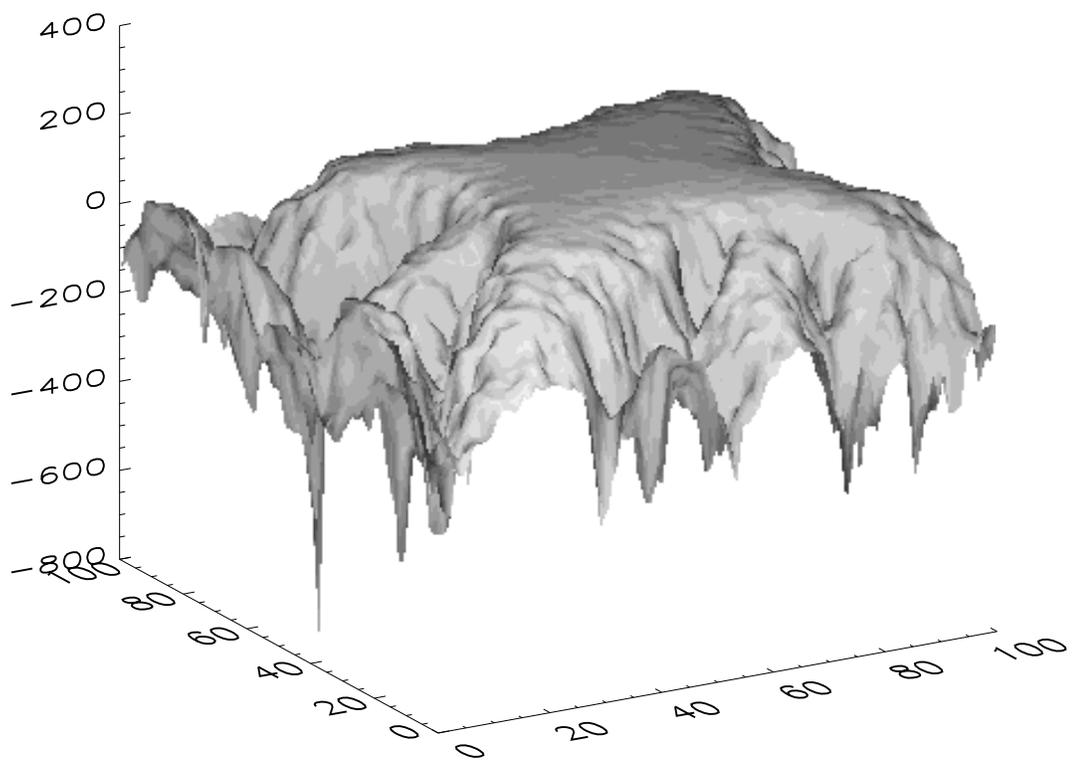
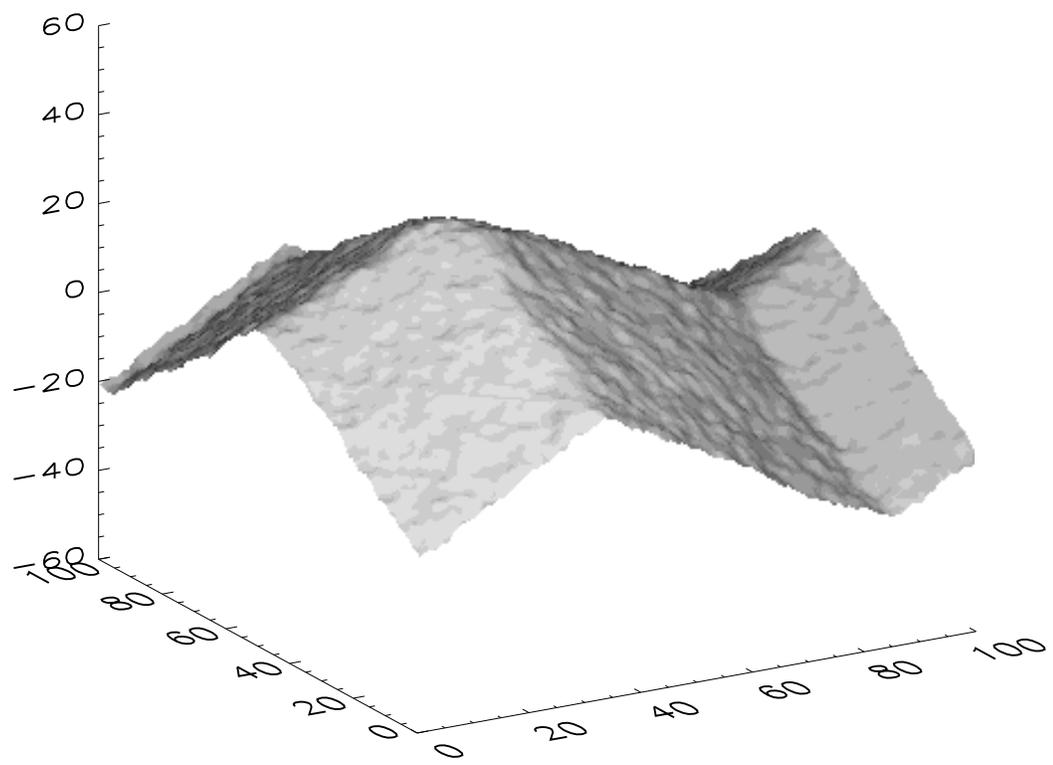
\end{document}